\documentclass[11pt]{article} 
\usepackage{aaspp4,psfig,epsf,rotate}
\slugcomment{Accepted to the Astrophysical Journal Letters}

\def\grb{GRB\,980329}

\def\etal{et al.}

\def\mjyb{~mJy~beam$^{-1}$}

\def\VLAJ{VLA~J070238.0+385044}

\lefthead{TAYLOR et al.}
\righthead{The Radio Afterglow from GRB\,980329}

\begin{document}

\title{The Discovery of the Radio Afterglow \\ From the Optically
  Dim $\gamma$-Ray Burst of March 29, 1998}

\author{G. B. Taylor\altaffilmark{1}, D. A. Frail\altaffilmark{1},
S. R. Kulkarni\altaffilmark{2}, D. S. Shepherd\altaffilmark{2},
M. Feroci\altaffilmark{3}, and F. Frontera\altaffilmark{4,5}}

\altaffiltext{1}{National Radio Astronomy Observatory, Socorro,
  NM 87801, USA} 

\altaffiltext{2}{Division of Physics, Mathematics and Astronomy 105-24,
  Caltech, Pasadena CA 91125, USA}

\altaffiltext{3}{Istituto di Astrofisica Spaziale, CNR,  
  via Fosso del Cavaliere, Roma I-00133, Italy}

\altaffiltext{4}{Istituto Tecn. Studio delle Rad. Extraterrestri,
  CNR, via Gobetti 101, Bologna I-40129, Italy}

\altaffiltext{5}{Dipartimento di Fisica, Universit\'a Ferrara, Via
  Paradiso 12, I-44100 Ferrara, Italy}

\begin{abstract}

  We report on the discovery of a variable radio source, \VLAJ,
  associated with the proposed x-ray counterpart, 1SAX J0702.6+3850 of
  \grb.  The source was monitored from one day after the burst to one
  month later at centimeter wavelengths (1.4, 4.9, 8.3 and 15 GHz;
  Very Large Array -- VLA), and in the 3-mm band (90 GHz; Owen Valley
  Radio Observatory -- OVRO).  Based on its variability, compactness
  and spectrum, we identify \VLAJ\ as the afterglow from \grb.  We
  interpret the rapid flux density variations as interstellar
  scintillation and the sharp turnover in the radio spectrum below 13
  GHz as arising from synchrotron self-absorption.  This suggests that
  the angular diameter of the fireball from \grb\ was of order a few
  microarcseconds in the first two weeks. The absence of a readily
  detectable afterglow in the optical, but clear detections in both
  the radio and infrared, can be understood as the result of
  extinction by dust. In this context it is interesting to note that
  half of all well-localized gamma-ray bursts have no optical
  counterpart. Based on our study of \grb\ we suggest that the
  progenitors of a sizable fraction of gamma-ray bursts are associated
  with regions of moderate to high gas density.

\end{abstract}


\vfill\eject
\section{Introduction}

The gamma-ray burst \grb\ was detected on 1998 March 29.1559 UT by the
gamma-ray burst monitor on the BeppoSAX satellite and localized to an
error circle of 3\arcmin\ radius by the Wide Field Camera (Frontera et
al.\ 1998). The burst lasted a total of 55 s and was bright -- a 
fluence of 5$\times{10}^{-5}$ erg cm$^{-2}$ in the range 50--300 keV
was measured by the Burst and Transient Experiment on board the 
Compton Gamma Ray Observatory
satellite (Briggs et al.\ 1998). Followup observations with the
Narrow Field X-ray Instruments (NFI) on board BeppoSAX began 7 hrs
after the burst.  in't Zand et al.\ (1998) detected a previously
uncataloged X-ray source 1SAX J0702.6+3850 (with an error circle of
1\arcmin\ radius) which was seen to fade by a factor of 3 over the 14
hr observation period. The burst position was further constrained by a
timing annulus derived between the Ulysses and BeppoSAX satellites
(Hurley et al.\ 1998).

A power law decay in the X-ray flux, S$_x\propto t^\delta$ appears to
be a universal signature of the X-ray afterglows; $\delta$ ranges from
$-1.1$ to $-1.5$ (e.g. Costa et al.\ 1997, Piro et al.\ 1998).  This
phenomenon of long-lived post-burst emission -- afterglow -- is nicely
explained by GRB models known as fireballs, in which relativistically
expanding debris shells interact with a surrounding medium and
accelerate a power law distribution of particles (M\'esz\'aros \& Rees
1997). As the fireball decelerates the peak of the emitted radiation
shifts in time to lower energies producing power law decays in the
X-ray, optical and radio bands (e.g.  Waxman 1997a, Sari, Piran \&
Narayan 1998).

The successful detections at X-ray wavelengths have not been matched
by equal numbers of afterglows at optical wavelengths. Only 4 of the
$\sim$10 GRBs with X-ray afterglows have detected optical afterglows
(van Paradijs et al.\ 1997, Bond et al.\ 1997, Halpern et al.  1998,
Groot et al.\ 1998).  GRB\,970828 is one notable non-detection; it
exhibited a typical X-ray decay (Murakami et al.\ 1998), yet no
optical transient was associated with this burst to a limit 100 times
lower than that of the optical afterglow from GRB\,970508 (Odewahn et al.
1997).  Several authors have attempted to explain the diversity of
afterglow behavior by invoking more complex fireball models
(M\'esz\'aros, Rees, \& Wijers 1998, Panaitescu, M\'esz\'aros, \& Rees
1998), including such effects as non-isotropic expansion, non-uniform
distributions of external media, and a time dependence in the radiative
efficiency of the shock.  Paczy\'nski (1998) suggested that foreground
extinction could account for the absence of optical afterglows from
GRBs. Indeed, modest extinction has been inferred for the optical
afterglow from GRB\,971214 (Reichart 1998, Halpern et al.  1998,
Ramaprakash et al.\ 1998). A logical extension of this extinction
hypothesis is that, in some cases, the afterglow may be seen in the
radio or infrared bands but not in the optical.

In this paper we report the detection of a variable radio source in
the reduced error circle of \grb\ and argue that it is the radio
afterglow from an optically obscured burst.

\section{Observations}


The first radio observations at 8 GHz with the Very Large Array
(VLA) began on 1998 March 30.212 UT.  Absolute flux calibration was
derived from short observations of 3C\,286 in the standard fashion, or by
using J0713+438 which also has a very stable flux density (Taylor,
Readhead \& Pearson 1996).  Phase calibration was derived from
observations of the nearby calibrators J0653+370 or J0713+438.


Continuum observations in the 3~mm band were made with the Owens
Valley Radio Observatory (OVRO) six element array between 1998 April 6
and 1998 April 11.  A total of four runs, each with approximately
6~hours of on-source time, were obtained under good conditions.
Projected baselines ranged from 30 to 120~m. Two 1~GHz bandwidth
continuum channels with central frequencies of 88.6 and 91.6 GHz were
observed simultaneously in the upper and lower side bands.
Flux calibration was derived using 3C\,273 with an absolute
uncertainty of $\sim 20$\%.  We used J0552$+$398 and J0646$+$448 for
phase calibration. 

\section{Results}

On 1998 April 1 (see Table 1), a modestly bright radio source, \VLAJ\ 
lying within the NFI error circle was detected.  The precise position
of this radio source is $\alpha$(J2000)~=~07$^{\rm h}$02$^{\rm
  m}$38\rlap{.}{$^ {\rm s}$}0217,
$\delta$(J2000)~=~38$^\circ$50$^{\prime}$44\rlap{.}{$^{\prime\prime}$}017
with an uncertainty of 0.05 arcsec in each coordinate.  Re-inspection
of the March 30 observation revealed a 3$\sigma$ detection at the
location of \VLAJ.  The flux density of the source dramatically
decreased on April 2 (Table 1). This clear variability led us to
announce that \VLAJ\ was potentially a radio afterglow of \grb\ 
(Taylor \etal\ 1998).

The source continued to exhibit variability as can be seen from Figure
1.  The modulation index (defined as the standard deviation divided by
the mean) for the period 1998 March 30 -- April 20 is 0.43 at 8.3 GHz
and 0.63 at 4.9 GHz. Beginning approximately three weeks after the
initial gamma-ray burst the modulation in the flux density decreased
(modulation index from April 21.91--30.10 is 0.30 and 0.39 at 8.3 and
4.9 GHz respectively).  
Perhaps the best evidence that the amplitude of the rapid flux
variations has subsided is seen in the distribution of spectral
indices in Table 1.  Between April 6 and April 21 $\alpha$ varies from
0.37 to 2.96, whereas after this time $\bar\alpha\simeq{2.6}$,
suggesting that the underlying intrinsic spectrum has been revealed.

\VLAJ\ is unresolved in all our observations.  The dataset of 1998
April 1 with its high signal-to-noise ratio allowed us to place
an upper limit on the angular size of \VLAJ\ of $<$50 milliarcseconds.
Based on observations on April 19.00 the linear polarization of \VLAJ\ 
is less than 21\% (2$\sigma$) at 8.3 GHz.

A summary of the OVRO results at 90~GHz is presented in Table 2.  In
the first two runs, there were marginal ($2\sigma$) detections of a
source within 2\arcsec\ of \VLAJ. The rms in the sum of the first
three days is 0.45 mJy ${\rm beam}^{-1}$. A source with a peak flux
of 1.45 mJy is found within 0.4\arcsec\ of \VLAJ.  


Smith \& Tilanus (1998) reported detections of a radio source
coincident with \VLAJ\ using the 15-m James Clerk Maxwell Telescope
(JCMT) at 350 GHz. Preliminary flux densities, on April 5.2, 6.2 and
7.2 UT were reported as 5$\pm$1.5 mJy, 4$\pm$1.2 mJy, and 2$\pm$0.8
mJy, respectively. Combining the first two nights Smith \& Tilanus
claim a mean flux density of 4.5$\pm$1 mJy. Within the stated
uncertainties, the JCMT source decayed very rapidly or the
measurements errors are larger than reported.  In view of this
somewhat confusing picture, we will restrict the discussion below to
measurements obtained from our VLA and OVRO efforts.

In Figure 2 we plot the time-averaged flux densities for the \VLAJ.  A
linear least squares fit gives a spectral index $\alpha\simeq$+0.9
(where $\alpha$ is defined by S$_\nu\propto\nu^\alpha$). Closer
inspection shows that while the radio spectrum rises with increasing
frequency, $\alpha$ flattens, with $\alpha$=1.7 between 4.9 and 8.3
GHz, and $\alpha$=0.8 from 15 GHz to 90 GHz.  We then fit the
predicted $\nu^{1/3}$ power-law spectrum (Waxman 1997b), attenuated by
a synchrotron self-absorption component of the form (1 $-$
e$^{-\tau}$)/$\tau$ where $\tau = \tau_\circ (\nu/\nu_\circ)^{-5/3}$.
As we discuss in \S{4.2}, Fig. 2 is sufficient to illustrate that a
$\nu^{1/3}$ spectrum is not inconsistent with the data at high
frequencies and that the synchrotron self-absorption turnover
frequency (where $\tau_\circ=1$) is close to 13 GHz.  The extrapolated
flux density at 350 GHz for this fit is 1.7 mJy.


What, then, is the nature of this compact, time-variable centimeter
radio source? The probability of finding an unrelated radio source at
8.3 GHz in the NFI error circle is small, 8\%, but it is not
negligible (Windhorst et al.\ 1993). However, the properties of \VLAJ\ 
are at odds with the majority of sub-milliJansky radio
sources at 5--10 GHz, which have flat spectral indices
($\bar\alpha=-0.4$), and are extended, ($\bar\theta$=2.6\arcsec)
(Windhorst et al.\ 1995, Richards et al.\ 
1998).  Optically they are usually identified with normal nearby
spirals or disturbed, faint blue galaxies (R$\simeq{22}$) whose radio
emission is thought to originate from recent bursts of star formation.

It is conceivable that \VLAJ\ could be a member of that small
fraction of the sub-mJy population that are identified with
low-luminosity active galactic nuclei (AGN).  The statistics of AGN
flux density variability are poor at these low flux density levels
(see summary in Frail 1998).  However, the variability for these sources is
broad-band and not the chromatic variability seen in Figure 1.

The properties of \VLAJ\ most closely resemble that of
the early-time behavior of the radio afterglow from the GRB of 1997
May 8 (Frail et al.\ 1997, Taylor et al.\ 1997). Like \VLAJ,
the radio source VLA~J065349.4+791619 was
initially undetected at 1.43 GHz, exhibited large flux variations at
4.86 and 8.46 GHz (Frail et al.\ 1997), and had a rising spectrum
toward higher frequencies (Shepherd et al.\ 1998). 
This marks only the second time a radio afterglow has been detected
from a GRB.  However, unlike the case for GRB\thinspace{970508}, no
clear optical signature was detected from the afterglow of \grb. In
the days prior to the report of the radio detection several groups
initially reported the absence of any optical variability in the
R-band in the range of R$>$20--23 (Guarnieri et al.\ 1998, Klose,
Meusinger, \& Lehmann 1998, Djorgovski et al.\ 1998a). Following our
announcement of the variable radio source (Taylor et al. 1998), we and
others re-examined the optical and IR data at the precise
sub-arcsecond position of the radio source.  We identified a faint
optical counterpart of \VLAJ\ (Djorgovski et al. 1998b).  Evidence for
a decaying counterpart was then found and reported as follows: I-band
(Klose 1998), J-band (Cole et al.\ 1998), R-band (Palazzi et al.\ 
1998a), and K-band (Larkin et al. 1998, Metzger 1998). These
observations provide ample confirmation that \VLAJ\ is the radio
afterglow from \grb.

\section{Discussion and Conclusion}

%

\subsection{The Mystery of Optically Dim GRBs}

In \S{1} we alluded to the fact that about half of the well-localized
GRBs appear not to have any associated optical transient.  There are
two solutions to this mystery (1) the afterglow decays rapidly in such
GRBs or (2) Many GRBs suffer from extinction.  Photoelectric
attenuation at keV energies decreases as $\nu^{-3}$ whereas optical
extinction $\propto \nu^q$ where $q$ is one or even larger.  Thus, a
modest column density, say $10^{21}$ cm$^{-2}$, in the host has little
affect on the SAX passband of 1--60 keV (Wide Field Camera and the
Narrow Field Instruments) but can suppress the observed R-band flux by
almost 5 mag.  Indeed the absence of detectable optical flux
accompanying strong X-ray emission may simply be due to the strong
redshift dependence of the suppression at observed optical wavelengths
in comparison to the X--ray wavelengths, $\tau_{\rm opt}/\tau_{\rm
  X-ray} \propto (1+z)^{3+q}$.

The standard afterglow model of M\'esz\'aros \& Rees (1997) predicts
that the peak flux density will be independent of frequency.  Our
average 8.3 GHz flux density of 325 $\mu$Jy from the period April
21--30 implies a peak R-band magnitude $\sim$ 17, yet only 15 hours
after the burst R was greater than 20 (Klose, Meusinger, \& Lehmann 1998).  
Observations of very red colors for \grb\ show that, at least for this
GRB, dust extinction is the cause of the dim optical emission
(Djorgovski et al. 1998c, Palazzi et al.\ 1998b).  Evidence for dust has been inferred in
GRB\,971214 (Reichart 1998, Halpern et al.\ 1998, Ramaprakash et al.\ 
1998); however, the inferred column density, while uncertain, is
rather modest, $A_R\sim 1$ mag.  In the case of \grb, the R-band
extinction is higher, making the detection of the afterglow easier at
IR wavelengths.

If dust is indeed the main cause of optically dim GRBs then we
conclude that extinction is significant in at least half of the GRBs.
At face value this is not compatible with the currently popular model
of coalescing neutron stars (Narayan, Paczy\'nski, \& Piran 1992). 
Such ns-ns binaries are expected to live long
enough ($10^9$ yr) and acquire sufficiently high velocities 
to remove them from their high gas density birthplaces.  
An alternative model for GRBs is the
``hypernova'' model (Woosley 1993, Paczy\'nski 1998), in which very
massive stars produce a ``dirty'' fireball with $\sim$300 times the
luminosity of a supernova.  Since such massive stars die young (age
$\sim10^6$ yr), a natural consequence of this model is that GRBs
should be associated with high density, dusty regions actively
undergoing star formation.  The large energies inferred for the
$\gamma$-ray burst and optical afterglow of GRB\,971214 (Kulkarni et
al.\ 1998; Ramaprakash et al.\ 1998) present a further challenge to
the coalescing ns-ns models. The hypernova model is capable of
producing a more energetic and long-lived afterglow via its ``dirty
fireball'' than current ns-ns models which produce a ``clean
fireball'' (Paczy\'nski 1998).

%
%
%

\subsection{Future Evolution of \VLAJ}

Taking cues from the radio afterglow of GRB\,970508 we expect that the
current chromatic intensity variations will die and be replaced by
broad-band lower modulation-index scintillation. Analyzing existing
and future data in terms of diffractive and refractive scintillation
will enable us to trace the angular evolution of the fireball (Frail
\etal\ 1997, Waxman, Kulkarni, \& Frail 1997). The character of the
fluctuations from \grb\ are strikingly similar to GRB\,970508 for
which a size of 3 $\mu$arcsec was inferred (Goodman 1997, Frail et
al.\ 1997). The angular size of \grb\ may be even smaller than this
since it is at a lower Galactic latitude ($b=18.7^\circ$ vs
26.7$^\circ$) and the quenching of scattering scales inversely with
sin($b$) (Goodman 1997).

When these flux modulations die down (due to the fireball expansion)
the underlying spectrum is revealed. Between 8.3, 15 and 90 GHz the
spectral slope $\alpha$ is roughly consistent with the canonical 1/3
value expected for the low energy tail of the electron distribution
(Fig.~2).  However, between 4.9 and 8.3 GHz a value of
$\alpha\simeq{2}$ is suggested at late times (see Table 1). A similar
slope was seen for GRB\thinspace{970508} between 1.43 GHz and 4.86
GHz (Shepherd et al.\ 1998), and was attributed to synchrotron
self-absorption (Katz \& Piran 1997). Again, as with GRB\,970508, we 
determine from this an angular size for \grb\ of a few $\mu$arcsec (Katz \&
Piran 1997). The higher self-absorption frequency inferred for \grb\ 
suggests that it is more compact than GRB\thinspace{970508}, or in
a denser environment, or that
there exists an additional component of low energy electrons (Waxman
1997b) that increase the opacity at radio wavelengths.  In keeping
with what has been observed from GRB\,970508 and model predictions,
the radio spectrum should evolve in time as the synchrotron peak
shifts to lower energies.  Within a few months we will have sufficient
data to carry out a comprehensive quantitative analysis of the radio
observations with respect to fireball models.


\acknowledgments

We thank E. Waxman and the referee, D. Helfand, for helpful suggestions.
The NRAO is a facility of the National Science Foundation operated
under cooperative agreement by Associated Universities, Inc. 
Research at the Owens Valley Radio Observatory is supported by the
National Science Foundation through NSF grant number AST 96-13717.  
SRK's research is supported by the National Science Foundation and 
NASA.

\clearpage

\begin{center}
TABLE 1 \\
\smallskip
VLA Observations of \grb
\smallskip
 
\begin{tabular}{l r r r r r r r r}
\hline
\hline
Epoch & $\Delta t$ & $S(4.9)$ & $S(8.3)$  &\multicolumn{1}{c}{$\alpha$}  \\
(UT 1998) & (d) & ($\mu$Jy) & ($\mu$Jy) & (4.9 -- 8.3)  \\
\hline
\noalign{\vskip2pt}
Mar.\ 30.23 & 1.07  &  -      &  166 $\pm$ 50  & - \\
Apr.\ 01.13 &  2.97 &    -    &   256 $\pm$ 16   & - \\
Apr.\ 02.15 & 3.99  &    -    &    84 $\pm$ 23   & - \\
Apr.\ 03.14$^a$ & 4.98  &  -    &   109 $\pm$ 47  & - \\
Apr.\ 04.11 & 5.95  &      -    &   135 $\pm$ 21  & - \\
Apr.\ 05.12$^b$ & 6.96  & -    &   194 $\pm$ 26  & -  \\
Apr.\ 06.89 & 8.73    &  22 $\pm$  37  &  58 $\pm$ 37  & 1.81 $\pm$ 1.80 \\
Apr.\ 08.06 & 9.90    & 146 $\pm$ 49  & 179 $\pm$ 41 & 0.37 $\pm$ 0.41 \\
Apr.\ 11.09 & 12.93    & 143 $\pm$ 45  & 274 $\pm$ 34  & 1.19 $\pm$ 0.34 \\
Apr.\ 18.04 & 19.88   &  107 $\pm$ 25 & 227 $\pm$ 22  & 1.41 $\pm$ 0.25 \\
Apr.\ 19.00 & 20.84   &   42 $\pm$ 21 & 205 $\pm$ 17  & 2.96 $\pm$ 0.51 \\
Apr.\ 20.27 & 22.11   &   52 $\pm$ 31 &  88 $\pm$ 21  & 0.98 $\pm$ 0.64 \\
Apr.\ 21.91 & 23.75   &   54 $\pm$ 20 & 291 $\pm$ 17  & 3.15 $\pm$ 0.37 \\
Apr.\ 23.10 & 24.94   &   65 $\pm$ 43 & 245 $\pm$ 37  & 2.48 $\pm$ 0.68 \\
Apr.\ 27.02$^c$ & 28.86   &  128 $\pm$ 42 & 465 $\pm$ 90 &  2.41 $\pm$ 0.38 \\
Apr.\ 30.10 & 31.94   &   85 $\pm$ 44 & 299 $\pm$ 51  & 2.35 $\pm$ 0.55 \\
\hline
\end{tabular}
\\
{\vskip2pt}
$^a$A 1.4 GHz flux density of 38 $\pm$ 45 $\mu$Jy was also measured.\\
$^b$A 1.4 GHz flux density of 48 $\pm$ 33 $\mu$Jy was also measured.\\
$^c$A 15 GHz flux density of 350 $\pm$ 130 $\mu$Jy was also measured.\\
\end{center}
\smallskip
\begin{flushleft}
{\sc Notes to Table 1:}
All observations were taken with the VLA
in the A configuration, providing a resolution of 1.5\arcsec,
0.5\arcsec, 0.25\arcsec, and 0.18\arcsec\ at 1.4, 4.9, 8.3, and 15 GHz
respectively. The total duration of each observing
run ranged between 0.5 and 5 hours.  The 1$\sigma$ errors for the
flux density measurements are given, which in general are proportional
to the square root of the time-on-source.  
\end{flushleft}
\bigskip
\clearpage
\begin{center}
TABLE 2 \\
\smallskip
OVRO Observational Summary
\smallskip
 
\begin{tabular}{lrrrcc}
\hline
\hline
Epoch &  S(90)         &$\alpha$(J2000) &$\delta$(J2000)  & \\
(UT 1998) & (mJy)         &(h m s)         &($^{\circ}~'~''$)&weather\\
\hline
Apr.\ 06 & 1.65 $\pm$ 0.70 & 07 02 38.03 & 38 50 44.77 & clear\\
Apr.\ 08 & 1.67 $\pm$ 0.85 & 07 02 37.85 & 38 50 44.28 & clear\\
Apr.\ 10 & 1.24 $\pm$ 0.79 & 07 02 37.96 & 38 50 44.34 &cloudy\\
Apr.\ 11 & 0.72 $\pm$ 0.80 & \nodata     & \nodata     &cloudy\\
First 3 runs \\
combined & 1.45 $\pm$ 0.45 & 07 02 38.02 & 38 50 44.39 &\nodata \\
\hline
\end{tabular}
\end{center}
\smallskip
\begin{flushleft}
{\sc Notes to Table 2:}
The images had a synthesized beam of $4.86''
\times 3.90''$ (FWHM) at position angle $-89.3^\circ$ with RMS noise
level $\sim 0.8${\mjyb} in a single run.
\end{flushleft}

\clearpage

 
\begin{figure}[htp]
\centerline{\psfig{figure=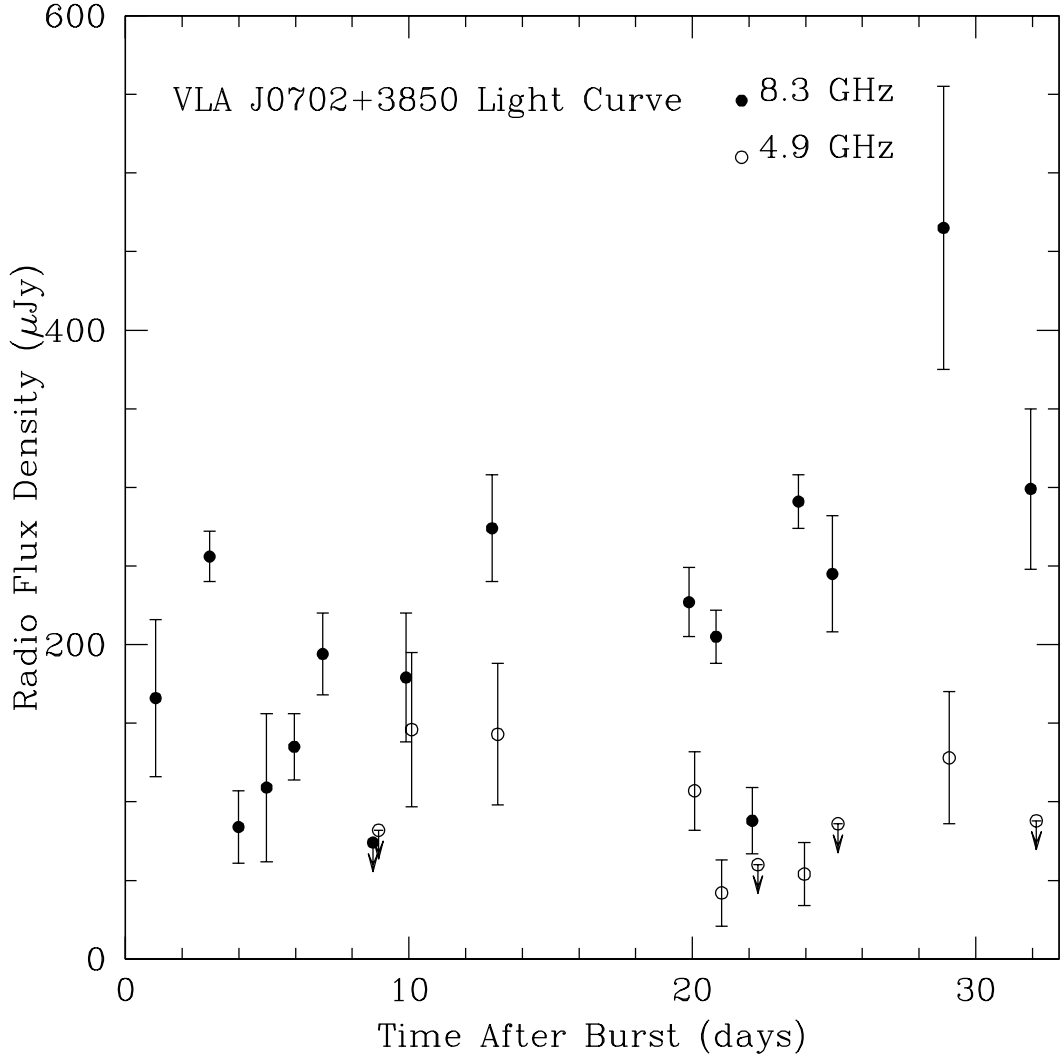,width=14cm,height=14cm}}
\figcaption{Radio light curve for \grb.  The time of the 4.9 GHz observations
have been offset by 0.2 days to avoid overlap with the 8.3 GHz 
measurements.  Upper limits (at 2$\sigma$) are shown for those
measurements with a significance of less than 2$\sigma$.}
\label{fig1}
\end{figure}
\clearpage

\begin{figure}[htp]
\centerline{\rotate[r]{\psfig{figure=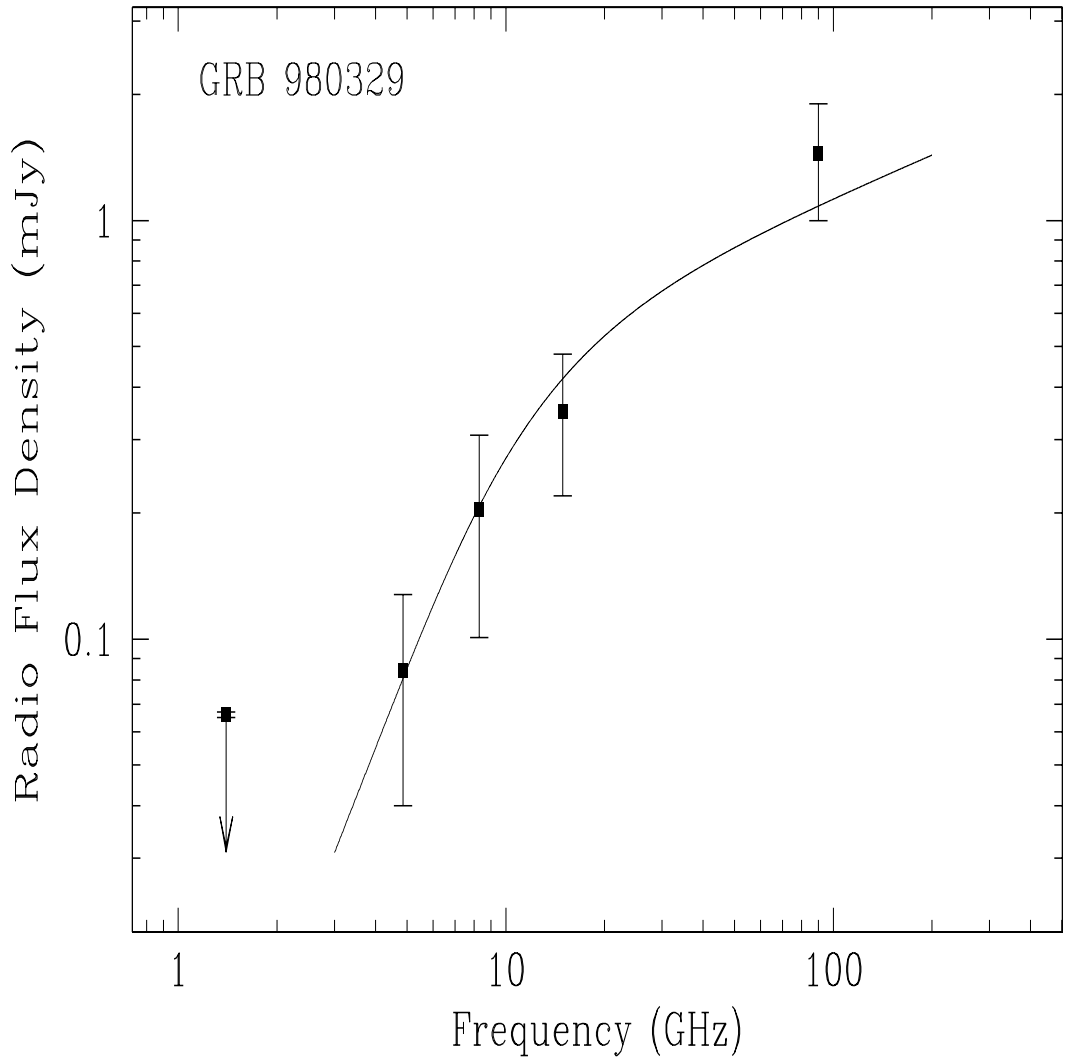,width=14cm,height=14cm}}}
\figcaption{The radio spectrum of \grb\ during the first month.  The
  average flux densities are plotted at 4.9 and 8.3 GHz.   The curve
shows a $\nu^{1/3}$ spectrum with a low frequency cutoff imposed
by synchrotron self-absorption.  The turnover frequency 
(where $\tau_\circ=1$) is $\sim$13 GHz.
}
\label{fig2}
\end{figure}



\clearpage
\end{document}